\documentclass[preprint,12pt]{elsarticle}

\usepackage{amssymb}
\usepackage{booktabs}

\usepackage{array} \newcommand{\PreserveBackslash}[1]{\let\temp=\\#1\let\\=\temp} \newcolumntype{C}[1]{>{\PreserveBackslash\centering}p{#1}} \newcolumntype{R}[1]{>{\PreserveBackslash\raggedleft}p{#1}} \newcolumntype{L}[1]{>{\PreserveBackslash\raggedright}p{#1}}

\begin{document}

\begin{frontmatter}



\title{Crystal growth, structure and physical properties of quasi-one-dimensional tellurides Fe$_{4-x}$VTe$_{4-y}$ ($x=1.01$, $y=0.74$) and V$_{4.64}$Te$_4$}


\author{Songnan Sun}
\author{Daye Xu}
\author{Chenglin Shang}
\author{Bingxian Shi}
\author{Jiale Huang}
\author{Xuejuan Gui}
\author{Zhongcen Sun}
\author{Juanjuan Liu}
\author{Jinchen Wang}
\author{Hongxia Zhang}
\author{Peng Cheng\corref{cor1}}

\address{Laboratory for Neutron Scattering and Key Laboratory of Quantum State Construction and Manipulation (Ministry of Education), School of Physics, Renmin University of China, Beijing 100872, China}

\cortext[cor1]{Corresponding author. \\E-mail
address:pcheng@ruc.edu.cn}

\begin{abstract}
A new ternary compound Fe$_{4-x}$VTe$_{4-y}$ ($x=1.01$,  $y=0.74$) with Ti$_5$Te$_4$-type structure is identified. Fe and V atoms tend to occupy different crystallographic positions and form quasi-one-dimensional (quasi-1D) Fe-V chains along the $c$-axis. Millimeter-sized single crystal of Fe$_{2.99}$VTe$_{3.26}$ (FVT) with slender-stick shape could be grown by chemical vapor transport method which reflects its quasi-1D crystal structure. Magnetization measurements reveal that FVT orders antiferromagnetically below T$_N$=93~K with strong easy $ab$-plane magnetic anisotropy. Although a weak glassy-like behavior appears below 10~K, FVT is dominant by long-range antiferromagnetic order in contrast to the spin-glass state in previously reported isostructural Fe$_5$Te$_4$. We also synthesize V$_{4.64}$Te$_4$ with similar quasi-1D V-chains and find it has weak anomalies at 144~K on both resistivity and susceptibility curves. However, no clear evidence is found for the development of magnetic or charge order. X-ray photoelectron spectroscopy and Curie-Weiss fit reveal that the effective moments for Fe$^{2+}$ and V$^{4+}$ in both compounds have large deviations from the conventional local moment model, which may possibly result from the formation of Fe/V metal-metal bondings. Furthermore the resistivity of both FVT and V$_{4.64}$Te$_4$ exhibits semiconducting-like temperature-dependent behavior but with average values close to typical bad metals, which resembles the transport behavior in the normal state of Fe-based superconductors. These quasi-1D compounds have shown interesting physical properties for future condensed matter physics research.   

\end{abstract}


\end{frontmatter}


\section{Introduction}
Transitional metal telluride has been a fertile playground for exploring materials with important physics such as unconventional superconductivity and low-dimensional quantum magnetism. For example, the parent compound of Fe-based superconductors FeTe, which can become superconducting either by chemical doping or in the form of thin-film\cite{1,2}. The discovery of gate-tunable room-temperature ferromagnetism in two-dimensional (2D) van der Waals Fe$_3$GeTe$_2$ has attracted enormous research interest in the Fe-Ge-Te 2D material family\cite{3}, which leads to the later observations of magnetic skyrmions and tunable magnetic ground states in them\cite{4,5,6,7}.

It has been well known that with reducing the dimension, plenty of novel quantum effects may emerge in low-dimensional materials. In quasi-one-dimensional (quasi-1D) conductors, electrons may condense into charge-density-wave (CDW) state due to Fermi surface instability\cite{8}. Furthermore, the electronic transport properties of 1D metals may violate the conventional Fermi-liquid theory and be described by the Tomonaga–Luttinger-liquid (TLL) theory\cite{9}. On the other hand, for magnetic materials with spin-$\frac{1}{2}$ 1D chains, exotic quantum states and fractionalized magnetic excitations such as spinons may be realized\cite{10}. Besides, many quasi-1D transitional metal tellurides are proposed to be 1D topological insulators with nontrivial topological states\cite{11,12,13}.
Therefore it would be worthwhile to explore new quasi-1D transitional metal tellurides, which may attract research interests from both chemists and physicists.

Binary compound Ti$_5$Te$_4$ was first discovered in 1961 and its crystal structure features quasi-1D Ti$_6$-octahedron chains along the $c$-axis\cite{14}. Thereafter, a lot of isostructural compounds were reported by replacing with other transitional metal (TM) and chalcogen elements. Some compounds with partial vacancy of the TM site such as Nb$_{4.7}$Te$_4$ and V$_{4.64}$Te$_4$ form different crystal structure but keep the main feature of quasi-1D TM$_6$-octahedron chains surrounded by Te atoms\cite{15}. However, so far as we know, the physical properties of all the above compounds were rarely investigated and whether their sizable single crystals could be grown is also unknown.  

In this study, we report the identification of a new quasi-1D transitional metal telluride Fe$_{4-x}$VTe$_{4-y}$ ($x=1.01$, $y=0.74$) with Ti$_5$Te$_4$-type structure. Through magnetization, transport and X-ray photoelectron spectroscopy measurements, it is identified as an antiferromagent with T$_N$=93~K in contrast to the magnetic states in Fe$_5$Te$_4$ and V$_{4.64}$Te$_4$ with similar TM$_6$-octahedron chains. Besides, both Fe$_{2.99}$VTe$_{3.26}$ and V$_{4.64}$Te$_4$ exhibit much reduced effective moments and resistivity behavior resembling the normal state of Fe-based superconducting family. These quasi-1D compounds have shown interesting physical properties that may attract future deep investigations.   

\section{Experimental details}

Single crystals of Fe$_{2.99}$VTe$_{3.26}$ (FVT) were grown by chemical vapor transport (CVT) method with iodine as the transport agent. The pure powder of Fe, V and Te were mixed in molar ratio of 4:1:4 and placed a quartz tube with certain amount of iodine. Then the evacuated quartz tube was sealed and put into a two-zone tube furnace. The quartz tube was heated to 750$\,^{\circ}\mathrm{C}$ in the raw material end and 700$\,^{\circ}\mathrm{C}$ in the other end in about 4 hours, then maintained at those temperatures for a week. Finally, the shining silver single crystals of FVT with typical size of 0.1~mm$\times$0.2~mm$\times$5~mm could be obtained in the cold end. 

Single crystals of V$_{4.64}$Te$_4$ could also be grown by similar CVT method. The initial mixture ratio for V:Te is 5.5:4 and the transport temperatures are set to be 50$\,^{\circ}\mathrm{C}$ higher than that of growing FVT for both ends. The obtained V$_{4.64}$Te$_4$ crystals were in the hot end with much smaller size than that of FVT (0.04~mm$\times$0.04~mm$\times$0.8~mm). It is difficult to perform physical property measurements on single crystals with such small size. Consequently, polycrystalline V$_{4.64}$Te$_4$ were synthesized by solid state reaction of V and Te powders with a molar ration of 5.5:4 at 1000$\,^{\circ}\mathrm{C}$ for 4 days. Then it is found that direct quench at this temperature is essential to obtain single phase of V$_{4.64}$Te$_4$.

Single crystal X-ray diffraction (XRD) patterns were collected at 273 K using a Bruker D8 VENTURE PHOTO II diffractometer equipped with multilayer mirror monochromatized Mo K$\alpha$ ($\lambda=$0.71073 \AA) radiation. Unit cell refinement and data merging were performed using the SAINT program\cite{16}. Structural solution was obtained for FVT and V$_{4.64}$Te$_4$ using the APEX3 program, and the final refinement was completed with the SHELXL suite of programs\cite{17}.

The powder samples were characterized by a Bruker D8 Advance X-ray diffractometer. The elemental composition of single crystals were examined with energy dispersive x-ray spectroscopy (EDS, Oxford X-Max 50), the results are consistent with the chemical formula. Magnetization and electrical transport measurements on bulk samples were carried out in Quantum Design MPMS3 and PPMS-14T respectively. The resistivity was measured with four-probe method.

X-ray photoelectron spectroscopy (XPS, Thermo Fisher K-Alpha+) was used to determine the valence state of the samples. The XPS spectrum was recorded using monochromatic Al K$\alpha$ X-ray source with a base pressure of 6.8×10$^{-9}$~Torr. Survey spectrum was recorded with a pass energy of 100~eV. The XPS data were calibrated by using adventitious C 1s signal at 284.8~eV as reference and the binding energy spectra were fitted by Avantage software. For the XPS depth profiling measurements, an Ar$^{+}$ energy of 1~keV at 10 mA ion current over a 2$\times$2~mm$^{2}$ area was used.

\section{Results and discussion}

\subsection{Crystal structure of FVT}

\begin{figure}
	\includegraphics[width=\textwidth]{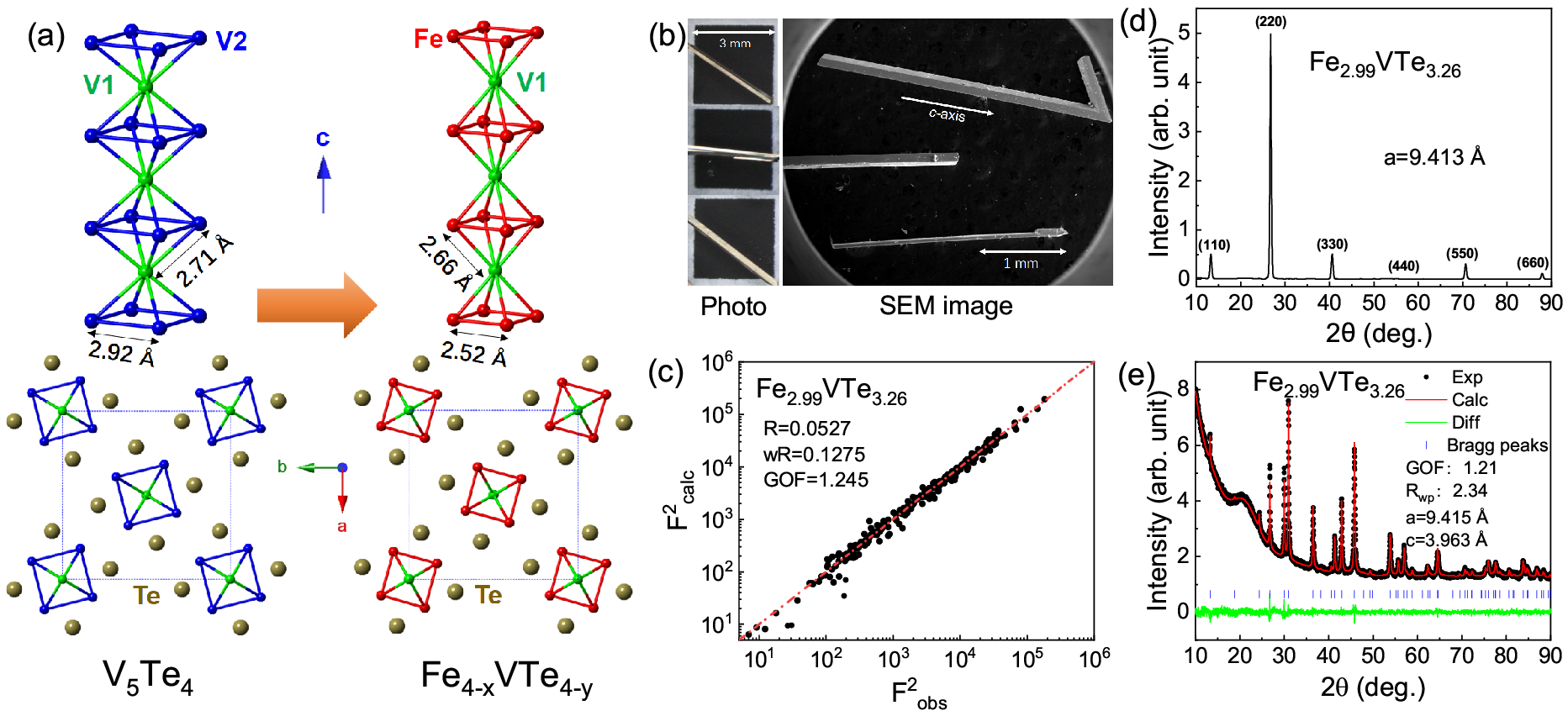}
	\caption {(a) Crystal structures of V$_5$Te$_4$ and FVT. The unit cell is
		shown with the blue dotted line. FVT can be viewed as a result of replacing all V2 atoms in V$_5$Te$_4$ by Fe atoms. (b) Optical photo and scanning electron microscope (SEM) image of FVT single crystals grown by CVT method. (c) The single-crystal X-ray refinement result on lattice Bragg peaks at room temperature. (d) Reflections on the shinning side surface of one FVT crystal. (e) Rietveld refinement result on FVT powders (crushed from single crystals).}\label{fig1}
\end{figure}

The crystal structure of FVT was initially solved through the refinement of single crystal XRD data. It adopts the tetragonal symmetry with space group $I/4m$ (No. 87), which belongs to the Ti$_5$Te$_4$-type structure. The obtained lattice parameters are $a=b=9.3992\AA$ and $c=3.9600\AA$. The refinement results on the lattice Bragg peaks are shown in Fig. 1(c) and more details are shown in Table S1 of the Supplemental Material. The structural features of FVT could be clearly seen through the comparison with V$_5$Te$_4$\cite{18} which is isostructural to Ti$_5$Te$_4$. The crystal structures of both FVT solved in this work and V$_5$Te$_4$ which is derived from the Inorganic Chemistry Structural Database (ICSD) are presented in Fig. 1(a). For V$_5$Te$_4$, there are two inequivalent Wyckoff positions for V atoms, namely V1 (2a) in the connection point of quasi-1D V$_6$-octahedron chains (green spheres) and V2 (8h) in other end points of the octahedron (blue spheres). Then FVT can be considered as a result of substituting Fe atoms for all the V2 atoms in V$_5$Te$_4$ while all V1 atoms remain the same (Table 1). This conclusion has been confirmed from both single crystal and powder XRD data [Fig. 1(e)]. Assuming the random occupancy of Fe/V on both $8h$ and $2a$ sites would lead the refined chemical formula significantly different from that determined by EDS. Although the XRD result supports a preferred occupancy of Fe/V on different sites, currently a slightly mixed occupancy of Fe/V on the same site cannot be rule out since these two ions have similar X-ray scattering length. 

Therefore the chemical formula of FVT with full occupancy should be Fe$_4$VTe$_4$. However, from both EDS measurement on the element composition and XRD refinement, it is found that there are notable vacancies on both Fe and Te sites. The refined values of occupancy on these sites are shown in Table 1 which are around 80\% and consistent with the values determined from EDS measurement. Any possible superlattice peaks related to superstructures or vacancy ordering are not observed. It should be mentioned that we have tried to tune the occupancy by varying the reagent ratio in sample preparation. Interestingly, all single crystal products have roughly similar occupancy as revealed from EDS and XRD measurement. Additionally, these crystals also have almost the same magnetic transition temperature and behavior which will be shown in the next subsection.

\begin{table*}
	\centering
	\caption{Atomic Coordinates, occupancy, and isotropic thermal parameters for F$_{2.99}$VTe$_{3.26}$.}
	\begin{tabular}{ccccccc}
		\toprule
		atom & Wyckoff & occupancy & x  & y & z & U(eq)  \\
        \midrule
		Te & 8h & 0.815 & 0.4275(1) & 0.7755(1) & 0.5 & 0.0168(6)  \\
		Fe & 8h & 0.747 & 0.3355(2) &  0.5938(2) & 0 & 0.0196(6) \\
		V & 2a & 1 & 0.5 & 0.5 & 0.5 & 0.0195(7) \\
		\bottomrule
	\end{tabular}
	\label{1}
\end{table*}

FVT features a quasi-1D Fe$_4$V$_2$-octahedron chains. Comparing with the V$_6$-octahedron chains in V$_5$Te$_4$, it is substantially elongated along the $c$-axis as shown in Fig. 1(a). This result agrees with the fact that FVT has a much larger $c$-lattice constant and a much smaller $a$-lattice constant comparing with that of V$_5$Te$_4$. The quasi-1D crystal structure of FVT is also manifested in the shape of single crystals which is slender-stick-like as shown in Fig. 1(b). The stick-like crystal has several shinning side surfaces. By collecting the XRD patterns from these surfaces, they are identified as (100), (110) or (310) planes. Reflections from (110) planes are shown in Fig. 1(d). These results further confirm that the long axis of the FVT crystal is the crystallographic $c$-axis.

\subsection{Magnetic and transport properties of FVT}

The magnetic properties of FVT are studied through anisotropic magnetization measurements on single crystals. Firstly, as shown in Fig. 2(a), a sharp drop of magnetic susceptibility $\chi(T)$ appears at T$_N$$\sim$93~K indicating the occurrence of an antiferromagnetic transition. This drop is much sharper for magnetic field along H$\parallel$$ab$ than that of along H$\parallel$$c$ [Fig. 2(b)], which serves as an evidence for an easy $ab$-plane magnetic anisotropy in FVT. We also find that there is not much difference for $\chi(T)$ data under field along different in-plane directions such as [110], [100] and [310]. This means a weak in-plane magnetic anisotropy and the direction of spins can be easily tuned within $ab$-plane, in contrast to in-plane anisotropic magnets\cite{19,20,21}. For $\mu_{0}$H=0.1~T, both the zero-field-cooling (ZFC) and field-cooling (FC) $\chi(T)$ data overlaps very well suggest the formation of long-range antiferromagnetic order. Then at below T$_g$$\sim$10~K, a small bifurcation between ZFC and FC data occurs which indicates an additional weak spin-glass-like component emerges.

\begin{figure}
	\includegraphics[width=\textwidth]{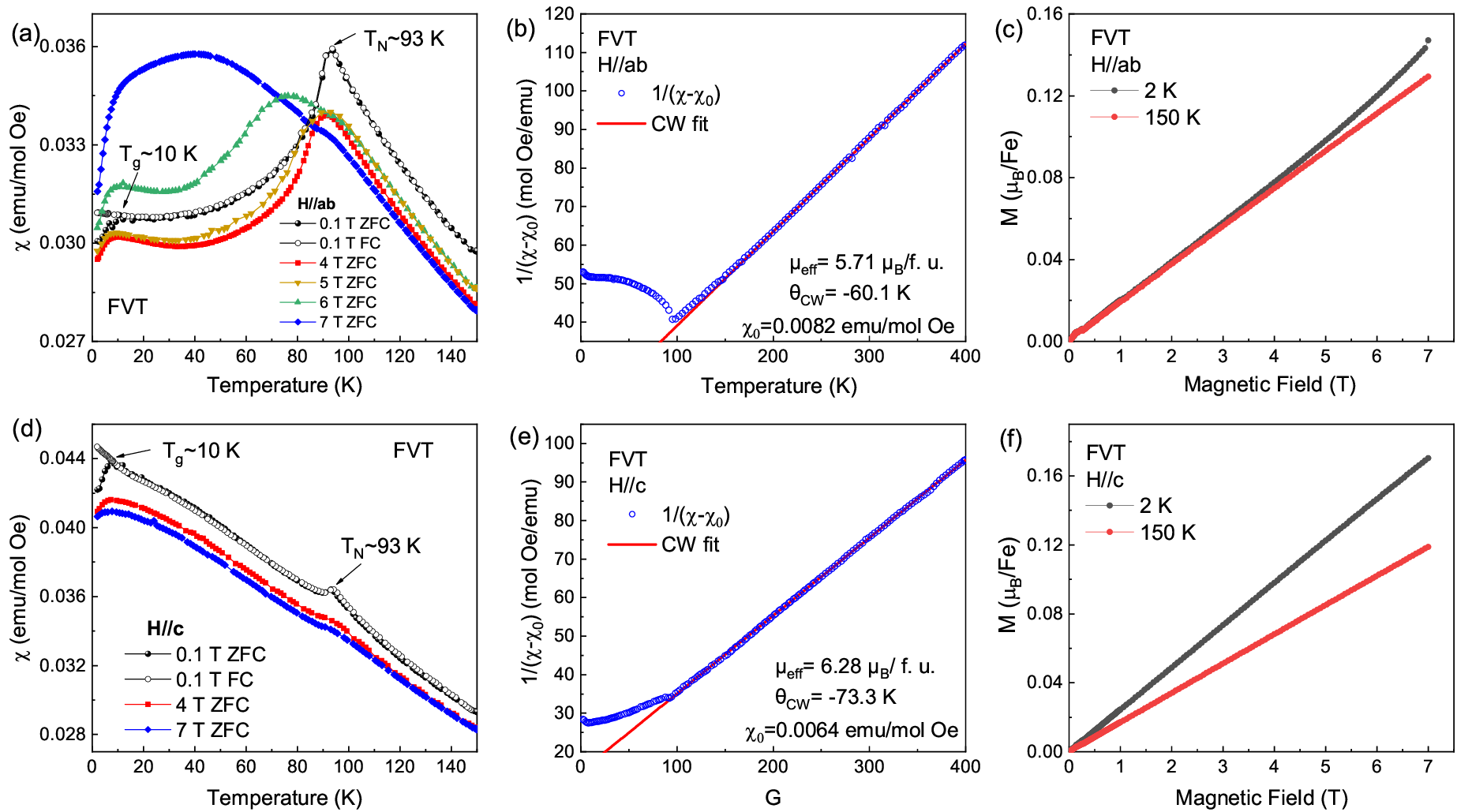}
	\caption {(a) Temperature dependent magnetic susceptibilities of FVT under magnetic field applied parallel to the $ab$-plane. (b) The result of Curie-Weiss fit on the high temperature paramagnetic magnetization data. (c) Isothermal magnetization under H$\parallel$$ab$ at two different temperatures. (d,e,f) The corresponding magnetization data of FVT single crystal under magnetic field applied parallel to the $c$-axis.} \label{fig2}
\end{figure}

With increasing field up to 7~T, the susceptibility anomaly caused by antiferromagnetic transition becomes weaker but still exists. Fig. S1 shows the d$\chi$/dT curve as a function of temperature which shows that T$_N$ slightly shifts to 89~K at 7~T. Furthermore, Fig. 2(a) shows the $\chi_{ab}(T)$ curves under field larger than 4~T are different from that below 4~T, namely a notable upturn feature emerges below T$_N$. This is a possible signature that a spin-flop transition makes the spins cant towards a perpendicular direction. As one can see from the isothermal magnetization data in Fig. 2(c) and (f), M(H) along H$\parallel$$ab$ at 2~K increases linearly with field below 4~T, but its slope increases quickly at higher fields which corresponds well with the expectation of a field-induced spin-flop transition. However, our instrument is unable to access fields higher than 7~T for a complete identification of this transition. Additionally, all other M(H) curves are linear with changing field, consistent with corresponding antiferromagnetic/paramagnetic states.

The inverse magnetic susceptibilities along different field directions are shown in Fig. 2(b) and (e). The data between 200~K and 400~K could be well fitted by the Curie-Weiss (CW) law combined with a temperature independent term $\chi_0$\cite{22}, which yields effective moment $\mu_{eff}=5.71~\mu_{B}/f. u.$ and CW temperature $\theta_{CW}$=-60.1~K for H$\parallel$$ab$, $\mu_{eff}=6.28~\mu_{B}/f. u.$ and $\theta_{CW}$=-73.3~K for H$\parallel$$c$. The negative $\theta_{CW}$ means dominant antiferromagnetic interaction in this compound and its absolute value is comparable with T$_N$. Interestingly, the effective moment per formula unit which contains four possible magnetic ions (three Fe and one V) is only around 6~$\mu_{B}$. This value is reproducible in several samples.

Fig. 3(a) displays the XPS spectra of Fe $2p$ core levels in FVT. The binding energies are identified as belonging to Fe $2p_{3/2}$, $2p_{1/2}$ and its satellite peaks, which are consistent with a Fe$^{2+}$ valence state. Similar analysis on XPS spectra of V $2p$ core levels shown in Fig. 3(b) reveal V$^{4+}$ valence state in FVT. Fig. 3(c) shows the local Fe-Te and V-Te bonds in FVT. Fe$^{2+}$ is surrounded by five Te ions and has a half octahedral crystal field environment. Therefore the $t_{2g}$ level should be lower and Fe$^{2+}$ may form either a high-spin S=2 (4.90~$\mu_{B}$/Fe$^{2+}$) or a low-spin S=0 state. For V$^{4+}$, it should have a S=1/2 (1.73~$\mu_{B}$/V$^{4+}$) state and the surrounded four Te ions form a square crystal field environment. It seems impossible for three Fe$^{2+}$ and one V$^{4+}$ in a formula unit to give a total effective moment close to the experimental value 6~$\mu_{B}$.

\begin{figure}
	\centering
	\includegraphics[width=\textwidth]{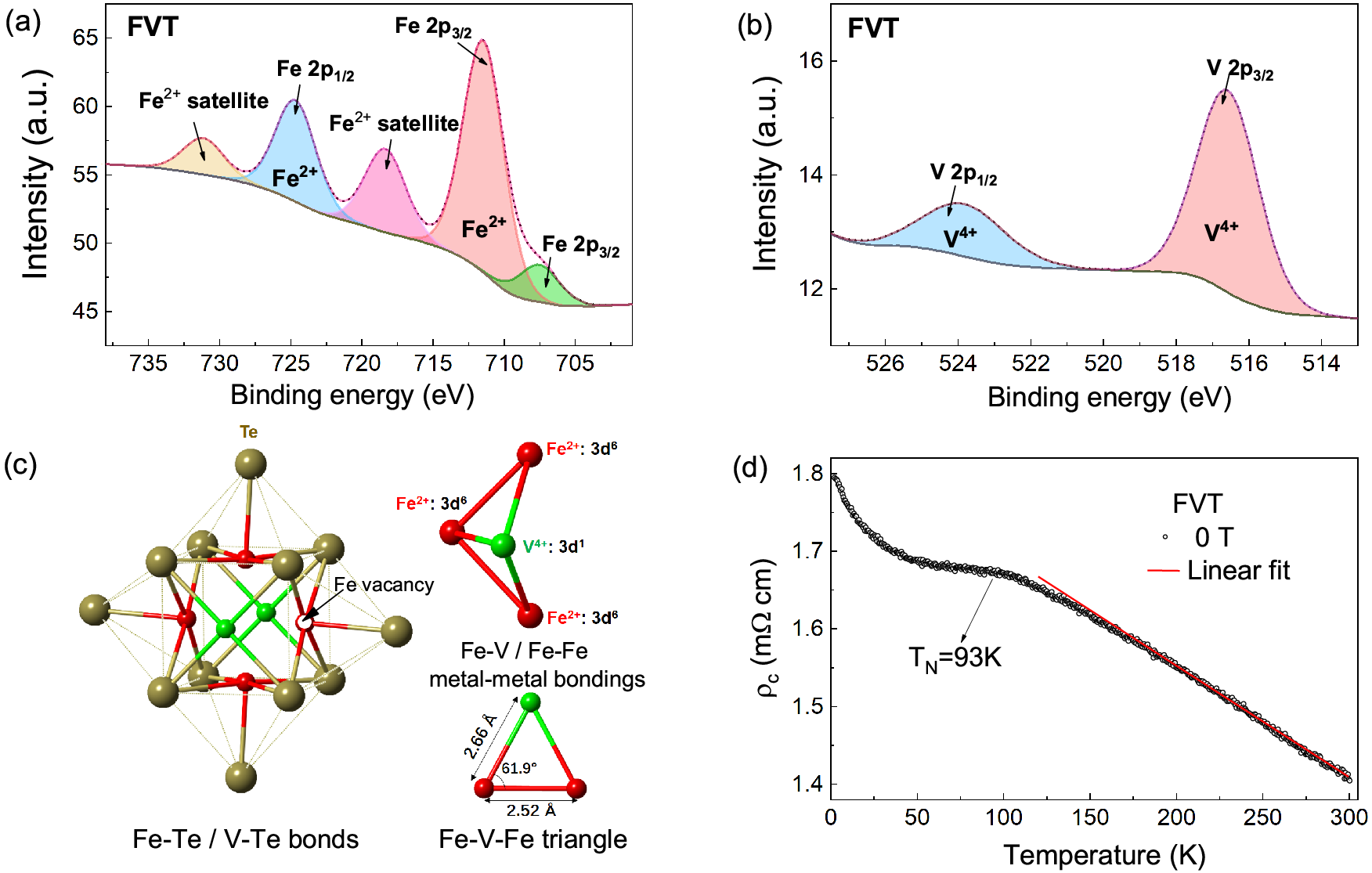}
	\caption {(a) XPS spectra of the Fe $2p$ peaks in FVT. (b) XPS spectra of V $2p$ peaks in FVT. (c) The local Fe-Te and V-Te bonds in a Fe$_4$V$_2$-octahedron, Fe-V and Fe-Fe metal-metal bondings in a formula unit and Fe-V-Fe triangles are illustrated. (d) Temperature-dependent electrical resistivity $\rho_c$ for FVT. The red solid line is a linear fit to the high temperature data.} \label{fig3}
\end{figure}

Next, we present a possible explanation to the above problem. Firstly, Fig. 1(a) and Fig. 3(c) show that the nearest intrachain Fe-Fe or Fe-V distances are quite short (below 2.7~\AA). Such a short distance means the formation of metal-metal bonding\cite{23}. For three Fe$^{2+}$ ions and one V$^{2+}$ ion in one formula unit, their metal-metal bondings may form molecular orbitals that influence the magnetism in this system. For example, in Nb$_3$Cl$_8$, three Nb ions form a Nb$_3$ trimer with $d_{Nb-Nb}$$\sim$2.8~\AA. As a result of the metal-metal bondings and formation of the molecular orbitals, the [Nb$_3$]$^{8+}$ clusters display S=1/2 magnetism\cite{23,24}. Similar novel magnetic states in Mo-based metal clusters were also reported\cite{25,26,27}. Then back to the situation of FVT, we speculate that the filling of nineteen $3d$ electrons from [Fe$_3$V]$^{10+}$ in the molecular orbitals formed by Fe-V-Fe metal-metal bondings generates an effective S=5/2 state with 5.92~$\mu_{B}$, which is close to the experimentally determined effective moment. The above picture is just an initial proposal and needs further theoretical investigations. The large deviation of effective moment from simple local moment picture has also been observed in many quantum materials such as cuprate and Fe-based superconductors with ongoing arguments on the underlying mechanism\cite{28,29,30,31}

The shape of FVT crystals makes the measurement of electrical transport along the $c$-axis more convenient and the $\rho_c(T)$ data is shown in Fig. 3(d). Below 300~K, the resistivity first increases linearly with decreasing temperature then a kink is observed at T$_N$=93~K, which demonstrates the impact of magnetic order on electrical transport. Although the temperature-dependent behavior of $\rho_c(T)$ is semiconducting-like, its absolute value is about $\sim$1.6~m$\Omega$~cm in average which is also in the range of a bad metal. Therefore, FVT may either be a semiconductor with very small band gap or a bad metal in which semiconducting temperature-dependent behavior could be driven by disorder effect. For the latter, Fe-based superconductors FeTe$_x$Se$_{1-x}$ could be served as similar examples. In previous reports\cite{1,32}, similar absolute value and temperature-dependent behavior for the normal state resistivity in FeTe, x=0.67 and x=0.75 samples are observed, which is explained as a disorder effect driven by the interplay between multiband effects and iron-$d$ shell band-filling.

\subsection{Structure and physical properties of V$_{4.64}$Te$_4$}

\begin{figure}
	\includegraphics[width=\textwidth]{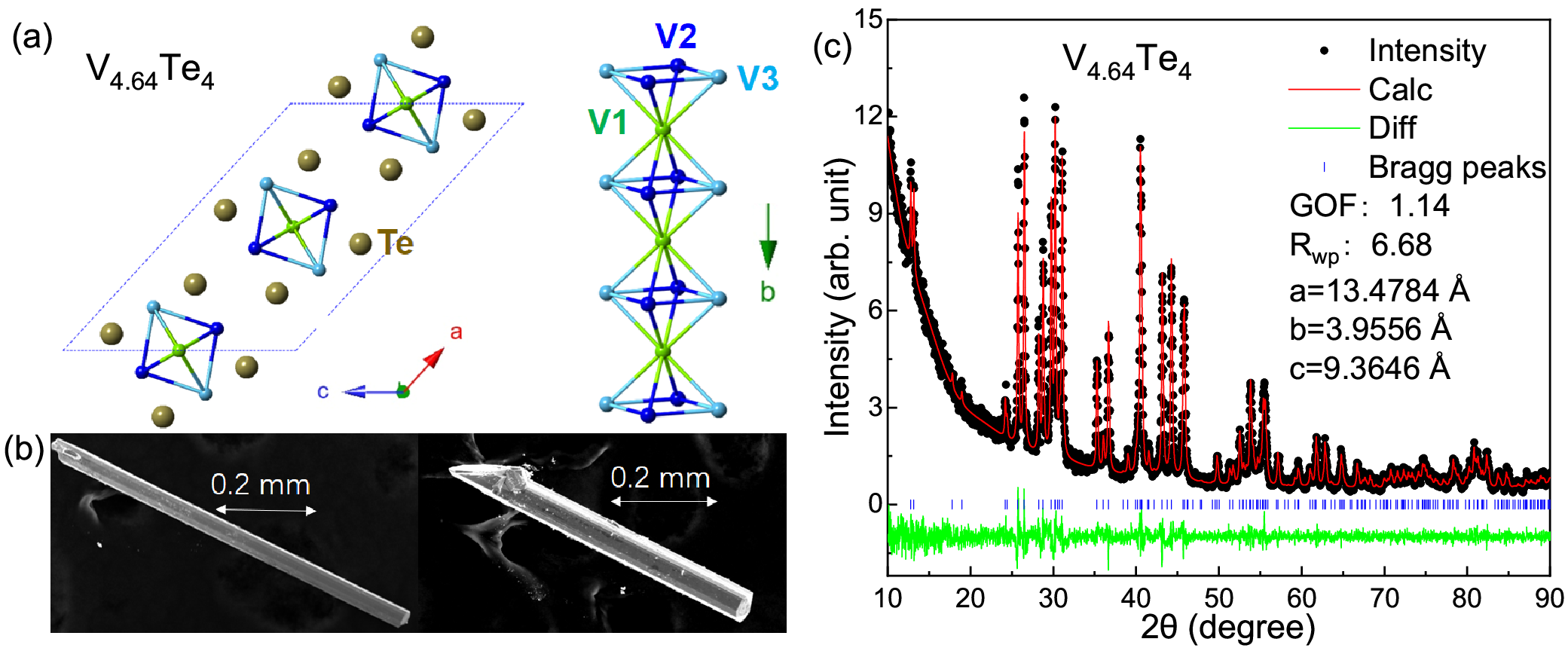}
	\caption {(a) Crystal structure of V$_{4.64}$Te$_4$. (b) SEM images of V$_{4.64}$Te$_4$ single crystals. (c) Room-temperature XRD patterns of polycrystalline V$_{4.64}$Te$_4$ and the Rietveld refinement result.}\label{fig4}
\end{figure}

Initially, we planned to synthesize V$_5$Te$_4$ with similar crystal structure for further investigations. However, by varying synthesizing conditions and reagents ratio, the products are all V$_{4.64}$Te$_4$\cite{15} as confrimed by XRD. So far as we know, the physical properties of both V$_5$Te$_4$ and V$_{4.64}$Te$_4$ are unknown in literature. According to the records in ICSD, V$_{4.64}$Te$_4$ crystallizes in the monoclinic space group $C12/m1$ (No.12)\cite{15}. It has quasi-1D V$_6$-octahedron chains surrounded by Te atoms along the $b$-axis as shown in Fig. 4(a) which is quite similar to that of V$_5$Te$_4$ and FVT. The reason for the change of space group is the small distortion of some V2 atoms. In V$_{4.64}$Te$_4$, V atoms have three different crystallographic positions marked by V1, V2 and V3 as shown in Fig. 4(a) and Table S3. The single crystal of V$_{4.64}$Te$_4$ has similar slender-stick shape as shown in Fig. 4(b), which reflects the quasi-1D like crystal structure. Analysis of single crystal XRD confirm its crystal structure is the same as that in previous report besides a slight difference in occupancy ratio and atomic coordinates\cite{15}. Detailed refinement and crystallographic data are presented in Table S2 and S3. The polycrystalline V$_{4.64}$Te$_4$ can also be directly prepared through solid state reaction and the powder XRD on these samples reveal consistent crystal structure as shown in Fig. 4(c).

Since the size of V$_{4.64}$Te$_4$ single crystals are too small, the magnetic and transport measurements were carried out on polycrystalline samples. Fig. 5(b) shows the temperature dependent magnetic susceptibility. $\chi(T)$ curve seems to display paramagnetic behavior down to 2~K, however a very weak anomaly could be found at around 144~K which could be easier to identify from the inverse susceptibility in the inset of Fig. 5(b). Considering a notable temperature-independent term $\chi_0$, the $\chi(T)$ data between 200~K and 400~K could be well fitted by the CW model, which gives $\mu_{eff}=0.93~\mu_{B}/f. u.$ and $\theta_{CW}$=-12.5~K. Since the indicated antiferromagnetic interaction is very weak, the 144~K anomaly is not likely a magnetic transition. The analysis on the XPS spectra of V$_{4.64}$Te$_4$ in Fig. 5(a) reveals a V$^{4+}$ valence state, while the obtained effective moment is only 0.20$\mu_{B}$/V$^{4+}$ which is much lower than a low-spin S=1/2 state. The reason might also be associated with the formation of V-V metal bondings shown in the inset of Fig. 5(a). Possibly in one formula unit, four $3d$ electrons from four V$^{4+}$ ions form S=0 singlets. The remaining 0.64 V$^{4+}$ with S=1/2 state leads to effective moment of 1.11~$\mu_{B}$ (0.64$\times$1.73~$\mu_{B}$), which agrees well with the effective moment 0.93~$\mu_{B}$/f. u. obtained from CW fit.

\begin{figure}
	\centering
	\includegraphics[width=13.5cm]{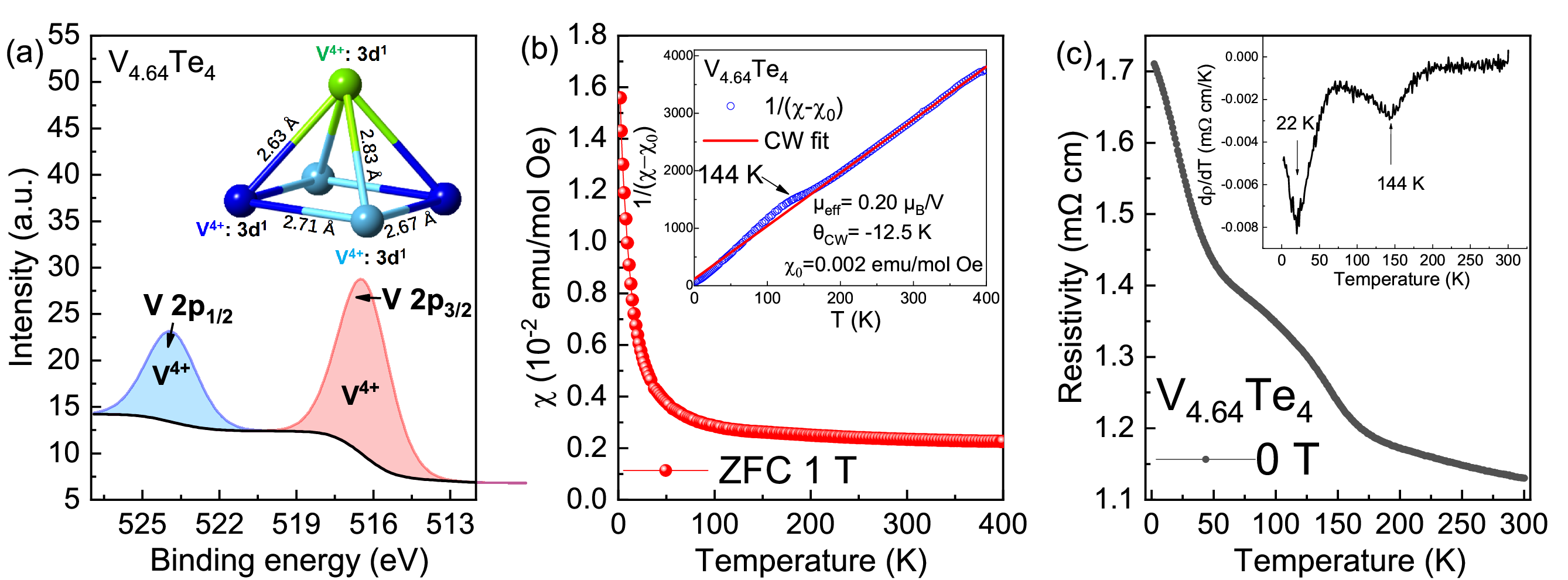}
	\caption {(a) XPS spectra of V $2p$ peaks in V$_{4.64}$Te$_4$. The inset shows the intrachain V-V bondings and distances. (b) Temperature dependent magnetic susceptibility of polycrystalline V$_{4.64}$Te$_4$. The inset shows the inverse magnetic susceptibility fitted by the Curie-Weiss law. (c) Temperature dependent resistivity of polycrystalline V$_{4.64}$Te$_4$ under zero field. The inset shows $d\rho/dT$ curve as a function of temperature.} \label{fig5}
\end{figure}

The result of the resistivity measurement is shown in Fig. 5(c). $\rho(T)$ also increases with decreasing temperature and has values similar to those of FVT, which indicates that V$_{4.64}$Te$_4$ is also a small-gap semiconductor or a bad metal. Interestingly, there are two resistivity anomalies at around 22~K and 144~K which are reproducible by measurements on samples from different batches. The latter anomaly is also identified in susceptibility curve. We performed single crystal XRD at 100~K for V$_{4.64}$Te$_4$, the results in Table S4 and S5 show that the structure remains the same and the lattice parameters do not have much difference. Moreover, no additional superlattice peaks could be observed in the precession photos of 100~K as shown in Fig. S2. Therefore, currently there is no evidence for the occurrence of possible charge-density-wave or magnetic transitions in V$_{4.64}$Te$_4$. Future investigations are needed to check whether there is any possible change of band structure which may be associated with the resistivity and susceptibility anomalies.

\subsection{Discussion}

FVT has very short nearest intrachain Fe-Fe distance (2.52\AA) and long interchain Fe-Fe distance (3.89\AA). Therefore one may expect quasi-1D magnetism in this compound as in SrCu$_2$O$_3$\cite{33} and CoNb$_2$O$_6$\cite{34}. However we have found that FVT has a long-range magnetic order with very high T$_N$. The magnetic susceptibility also does not follow Bonner Fischer's law. These facts rule out the possibility for quasi-1D magnetism and mean that the interchain magnetic interaction may also be strong. The magnetic superexchange interaction between Fe ions mediated by Te ions possibly exists.

On the other hand, the development of long-range magnetic order in FVT is a surprising experimental fact comparing with the short-range spin-glass state in isostructural Fe$_5$Te$_4$. Previously Ulbrich $et~al.$ reports an iron-rich telluride Fe$_5$Te$_4$\cite{35} phase. Fe$_5$Te$_4$ has the same crystal structure as FVT and can be viewed as V-doped Fe$_5$Te$_4$. Ulbrich $et~al.$ found that Fe$_5$Te$_4$ exhibits a magnetically frustrated spin-glass state below 150 K, due to the competing ferro-/antiferromagnetic interactions. It is important to mention that usually chemical doping or atomic disorder will drive a long-range magnetically ordered system to be a short-range spin-glass system\cite{5,36,37,38,39}. Comparing with Fe$_5$Te$_4$, FVT should be a more disordered system with V-doping and additional disorder from Fe/Te vacancies, which should hinder the formation of long-range order. Remarkably, the magnetic frustrated spin-glass in Fe$_5$Te$_4$ is actually suppressed in FVT with the development of dominant long-range antiferromagnetic order. The frustrated spin-glass in Fe$_5$Te$_4$ possibly arises from the Fe$_3$ triangles which constitute the Fe$_6$-octahedral chains. Since Fe$_5$Te$_4$ has the same crystal structure as FVT, one can see this Fe$_3$ triangle from Fig. 3(c) imagining the green V atom is Fe atom. Although not strictly equilateral, it may create some frustration in Fe$_5$Te$_4$. Then replacing Fe in the $2a$ Wyckoff position by V naturally modify the magnetic interaction from Fe-Fe direct or Fe-Te-Fe superexchange interactions to Fe-V or Fe-Te-V exchange interactions, which may suppress the magnetic frustration and lead to long-range order. The magnetic state formed by Fe-V-Fe metal-metal bondings mentioned before may also favor the development of long-range magnetic order.

Finally, the physical properties of compounds with Ti$_5$Te$_4$-type structure have been rarely reported. Our current results may stimulate further researches in this material family, such as tuning these compounds by pressure or chemical doping to explore possible superconductivity or charge-density-wave\cite{40,41}. As in K$_2$Cr$_3$As$_3$ with quasi-1D Cr-chains and metal-metal bondings, unconventional superconductivity has been discovered\cite{41}.

\section{Conclusions}
In summary, we have successfully synthesized a new ternary compound F$_{2.99}$VTe$_{3.26}$ with Ti$_5$Te$_4$-type quasi-1D crystal structure. FVT is an antiferromagnet with Neel temperature of 93~K and an easy $ab$-plane magnetic anisotropy. This dominant long-range magnetic order is in sharp contrast to the short-range spin-glass state in isostructural Fe$_5$Te$_4$. V$_{4.64}$Te$_4$ with structural similarities is also prepared. Its resistivity and magnetic susceptibility have an anomaly at 144~K. For both FVT and V$_{4.64}$Te$_4$, the values of effective moment per formula unit obtained from CW fit are different from the expectation based on conventional local moment picture, which might be due to metal-metal bondings in quasi-1D TM$_6$-octahedron chains. Besides, their resistivities have semiconducting temperature-dependent behavior but with absolute values close to a bad metal, resembling the normal state transport behavior in Fe-based superconductors. The above results may attract further research interest in these quasi-1D compounds.

\section*{Acknowledgement}
We appreciate the useful discussions with Prof. Yanfeng Guo. This work was supported by the National Natural Science Foundation of China (No. 12074426, No. 12474148).

\label{}





\end{document}